\documentclass{emulateapj}
\usepackage{natbib}
\usepackage{graphicx}
\usepackage{longtable}
\usepackage{url}
\begin{document}
\providecommand{\e}[1]{\ensuremath{\times 10^{#1}}}

\submitted{Accepted to PASP}
\title{Future Direct Spectroscopic Detection of Hot Jupiters with IGRINS}
\author{Kevin Gullikson\altaffilmark{1},
             Michael Endl\altaffilmark{1}}

\altaffiltext{1}{Astronomy Department, University of Texas, 1 University
Station C1400, Austin, TX 78712, USA. \texttt{kgulliks@astro.as.utexas.edu}}

\begin{abstract}
With about 700 confirmed extrasolar planets, it is time to move beyond discovery and towards characterization. Perhaps the most basic parameter of an extrasolar planet is its mass; however, this is very difficult to determine if the planet does not transit the star. The radial velocity technique, still the most fruitful method of discovering planets in the solar neighborhood, can only determine a minimum planet mass. We investigate a method using the near-future IGRINS near infrared spectrograph to detect the orbital motion of the planet itself. We simulate several observations of a star with an orbiting planet, and search for the spectral signature of the planet by cross-correlating against planet model spectra. A detection appears as a strong peak in the cross-correlation function, and gives the radial velocity of the planet at the time of observation. This, combined with the motion of the star from traditional radial velocity planet search programs, can determine the actual planet mass. We find that the IGRINS instrument can detect the spectral signature from large planets on very close orbits (so-called Hot Jupiters), and that the detections can provide tight constraints on the true planet mass.
\end{abstract}


\maketitle

\section{Introduction}
With about 700 confirmed extrasolar planets \citep[from
exoplanets.org:][]{exoplanet}, the time for characterization of these
planets is here. A first step towards characterization is a determination of
the planet mass. Most of the exoplanets so far discovered around nearby stars were
found using the radial velocity technique, which measures the periodic
Doppler shift of the parent star. Unfortunately, the inclination of the orbit cannot be determined
without another complementary method. This means that planet masses from
radial-velocity surveys are only \emph{minimum} masses. In principle, precise astrometry could provide the complementary measurement that is needed to determine the true planet mass. However, the astrometric motion is very difficult to detect with current technology. The amplitude of the motion of a sun-like star with a Jupiter-mass planet orbiting at 0.1 AU and a distance from Earth of 10 pc is roughly $20 \mu$as, an order of magnitude below the precision of the Hubble Space Telescope Fine Guidance Sensors \citep{Benedict2006} and will be very near the precision of GAIA \citep{Sozzetti2001}.

The true mass and inclination of a planet could be determined if the
radial velocity of the planet was known as well as that of its parent
star. There are two ways that the
radial velocity of an orbiting planet could be measured: light
reflected from the parent star or the characteristic spectrum of the planet itself. Both methods require high resolution spectroscopy in order to detect the doppler motion of the spectral lines. Several groups have
attempted to detect the reflected light from orbiting planets, but at
the time of this writing none have been successful \citep{Collier2002,
  Rodler2008, Rodler2010, Langford2011} and have only been able to set upper limits on the planet albedo ($A_B \sim 0.1
$).

While searches for reflected light are best done in the optical, the
thermal emission from a $\sim 1000$K planet will peak in the
near-infrared. The thermal emission from a small group of planets has been detected,
including HD209458b \citep[e.g.][]{Knutson2007, Swain2008,
  Cubillos2010}, HD189733b \citep[e.g.][]{Grillmair2007, Knutson2007_2,
  Char2008, Agol2010}, Wasp-3b \citep{Zhao2012}, and even the
Super-Earth 55 Cnc b \citep{Demory2012}. These detections were mostly made
using either Spitzer photometry or low-resolution spectroscopy, and
they are all transiting planets. Recently, the emission spectrum from Tau Boo b \citep{Brogi2012, Rodler2012}, HD 189733b \citep{deKok2013}, and possibly 51 Peg b \citep{Brogi2013} has been detected in high resolution using VLT/CRIRES. In this paper, we describe a similar technique using the IGRINS instrument, which offers a much larger spectral range in a single observation than CRIRES, and is expected to see first light on the 2.7m Harlan J. Smith Telescope at McDonald Observatory in late 2013.

There are several challenges to detecting the planet's near-infrared spectrum, especially in high resolution. The very low flux ratio between the planet and the star ($F_p/F_s \sim 10^{-4}$ in the K-band) requires a very sensitive instrument and a high signal-to-noise ratio (S/N), which is very challenging on current near-infrared spectrographs. Second, the near-infrared spectrum is highly contaminated by absorption from the Earth's atmosphere (telluric absorption). In order to detect a planetary spectrum, the telluric lines must be removed very well. Finally, the stellar spectrum must be removed to detect the planetary spectrum. This is extremely challenging for non-transiting planets, for which the planet is never blocked by the star. 

In this paper, we investigate a technique to detect the spectrum from an approximately Jupiter mass object on a very close orbit (a Hot Jupiter) using the near-infrared spectrograph IGRINS. We briefly describe the IGRINS instrument and the simulated observations in Section \ref{sec:method}. In Section \ref{sec:results}, we test the sensitivity of our detection method to the S/N of the observations, the efficiency of heat redistribution from dayside to nightside in the planet's atmosphere, and the model atmosphere dependence of the method. We summarize our results and compare our sensitivity to the recent detections of Hot Jupiters in Section \ref{sec:summary}.

\section{Instrument and Methodology}
\label{sec:method}
The IGRINS instrument is explained in detail in \cite{IGRINS}. IGRINS
is an immersion grating echelle spectrograph, and will
be capable of observing the entire H (1.4-1.8 $\mu m$) and K (2.0-2.4 $\mu m$) spectral windows at once,
with a resolution of $R=\lambda / \Delta \lambda = 40000$. IGRINS is
scheduled to see first light on the 2.7-meter Harlan J Smith telescope at McDonald
Observatory in late 2013.

We make several assumptions and simplifications in this work. We
ignore any \emph{instrumental} effects that may introduce systematic noise, although we do introduce systematic noise in the form of telluric contamination. We also ignore any light reflected from the star, since it contributes little to the total planet brightness in the H and K bands. Finally, we assume the IGRINS detector has equal sensitivity to all wavelengths of light, so the signal-to-noise ratio is set only by the light from the target.

There are three main steps in our simulated observing program: generating
a series of synthetic observations, removing the signature of Earth's
atmosphere as well as the parent star's spectrum, and searching for
the planet signal in the residuals. Each of these steps is detailed
below.

\subsection{Synthetic Observation Generation}
\label{sec:obsgen}
An observed spectrum of a star and planet system can be divided into
three parts: the star, the planet, and the Earth's atmosphere
(telluric contamination). We use two test cases for this work: HD 189733 and HD 209458. The basic parameters of the systems are given in Table \ref{tab:planet}. We choose these planets as test cases becase they are very well-studied, with detections of the atmosphere both in transmission and emission. As a result of this, they are one of the very few planets with constrained atmospheric temperature-pressure profiles and chemical abundances. These planets are also useful as test cases since HD 209458b is thought to have a thermal inversion layer in its atmosphere \citep{Knutson2008}, while HD 189733b does not \citep{Char2008}.

\begin{table*}[t]
  \centering
  \begin{tabular}{|c|ccccc|}
    \hline
    Star & $T_{\rm eff, star}$ (K) & $T_{\rm eff, planet}$ (K) & M$_{\rm p}$ (M$_{\rm Jup}$) & R$_{\rm p}$ (R$_{\rm Jup}$)  & $r$ \\ \hline
    HD 189733 & 5040 & 1200 & 1.14 & 1.138 & 0.77 \\
    HD 209458 & 6065 & 1450 & 0.69 & 1.359 & 0.90 \\ \hline
  \end{tabular}
  \caption{Physical properties of the HD 189733 and HD 209458 planetary systems. The effective temperatures, masses, and radii are from \cite{Torres2008}, and r (the ratio of the flux in an average spectral line to the flux in the continuum for the planet) are from the model spectra used in this work (See Figure \ref{fig:modelcomp} and the discussion before Equation \ref{eqn:snrcrit}).}
  \label{tab:planet}
\end{table*}

 We simulate the stellar and planetary spectra using a code based
on the Phoenix-ACES stellar atmosphere code \citep[described in][]{Barman2011}, modified to self-consistently treat a planet with intense incoming stellar radiation by using the stellar radiation field as a boundary condition on $F_{\nu}$.  \citep{Barman2001}. For this work, we used one dimensional spherical geometry, with no clouds and solar abundance ratios \citep{Asplund2005}. The temperature-pressure profiles are described in \cite{Barman2008} and \cite{Barman2002} for HD 189733b and HD 209458b (respectively). The chemical abundances are solved at each layer by assuming complete chemical equilibrium.

The absorption due to earth's atmosphere was modeled using the Line-By-Line Radiative Transfer Model (LBLRTM)
code \citep{Clough2005}. This code takes the pressure, temperature,
and abundance of several molecular species at a series of heights in
the atmosphere, and outputs a transmission spectrum. The code also
requires a line list containing the molecular
line strengths and positions which, along with the molecular abundance and the airmass of the observation,
determines the amount of absorption at a given wavelength. We use the HITRAN database \citep{Rothman2009} for the line
list.

The synthetic observations were made by first adding the star and
planet model spectra at the appropriate flux ratio and
Doppler shifts. The flux ratio was determined by the model
spectra themselves, multiplied by the radius of the body. We determined the Doppler shift by fixing the orbital parameters and masses of both the star and planet (which we will ultimately recover). Synthetic observations were made at
several phases in the planet's orbit as well as different phases of the \emph{Earth's} orbit around the Sun, resulting in a variety of relative radial velocities between the star, planet, and telluric spectral lines. 

Hot Jupiters are expected to be tidally locked with their parent stars
\citep{Fabrycky2010}, meaning there are permanent day and night sides. The extent
of heat redistribution from the dayside to the nightside is uncertain, 
but appears to vary throughout
the Hot Jupiter planetary class. \cite{Knutson2007_2} find that
HD189733 b is consistent with a high degree of heat redistribution
between its day and night side. Conversely, \cite{Harrington2006} find
that $\nu$ And b is consistent with no heat redistribution. 

For planets with little heat redistribution, the planetary spectrum will change throughout the orbit, as different amounts of the cold nightside are exposed. Without a full suite of planetary atmospheres calculated with a self-consistent phase curve, we cannot treat the issue of heat redistribution in a fully realistic way. We approximate a planet with inefficient heat redistribution by scaling the dayside planet spectrum with a phase-dependent factor before adding it to the stellar spectrum. In this work, we consider only two cases: one with complete heat redistribution where the temperature of the planet is invariant to the orbital phase, and a second where the effective temperature seen on the nightside is some fraction ($f_{\rm red}$) of the dayside temperature, resulting in a minimum scaling factor of $f_{\rm red}^4$. WASP-12b, a very extreme case, has $f_{\rm red} = 1/3$ \citep{Cowan2012};  we simulate observations with $f_{\rm red} = 1/2$ as a reasonable value, and adopt a sinusoidal phase curve. For this case, the nightside planet spectrum is $2^4 = 16$ times dimmer than the dayside spectrum.

After adding the star and planet spectra as above, we multiply the sum by a model of the telluric absorption spectrum. We then convolve the spectrum with a gaussian instrumental profile and rebin the data according to the predicted resolution and spectral format of IGRINS \citep[Figures 2 and 3 of ][]{IGRINS}. Finally, we set the average signal-to-noise ratio by adding gaussian random noise to each pixel.

\subsection{Telluric and Stellar Line Removal}
\label{sec:tellcorr}
We now begin attempting to recover the planet spectrum from the synthetic observations. To remove the
telluric contamination, we use a method similar to that described by
\cite{Rodler2012}. With this method, a ``telluric standard'' star, which is usually an A- or B-star, is observed immediately after the science target at a similar airmass. The telluric contamination is modeled independently for both stars, and then the residuals of the science star fit are divided by the residuals of the standard star fit. Using a telluric model accounts for changes in airmass, water vapor column density, and the instrumental profile between the science and standard stars, but can leave systematic errors if certain lines have incorrect oscillator strengths. Dividing the residuals after the telluric model fit removes most of these systematic errors. We simulate observations of the science target
as described above, and simulate the observation of a B-type telluric standard
star with a 20000 K blackbody. The generation of this synthetic observation
was identical to that used for the science star, except we used a
telluric model with a different telescope altitude (airmass) and we did not add the planet model spectrum. 

\begin{figure}[t]
  \centering
  \includegraphics[width=\columnwidth]{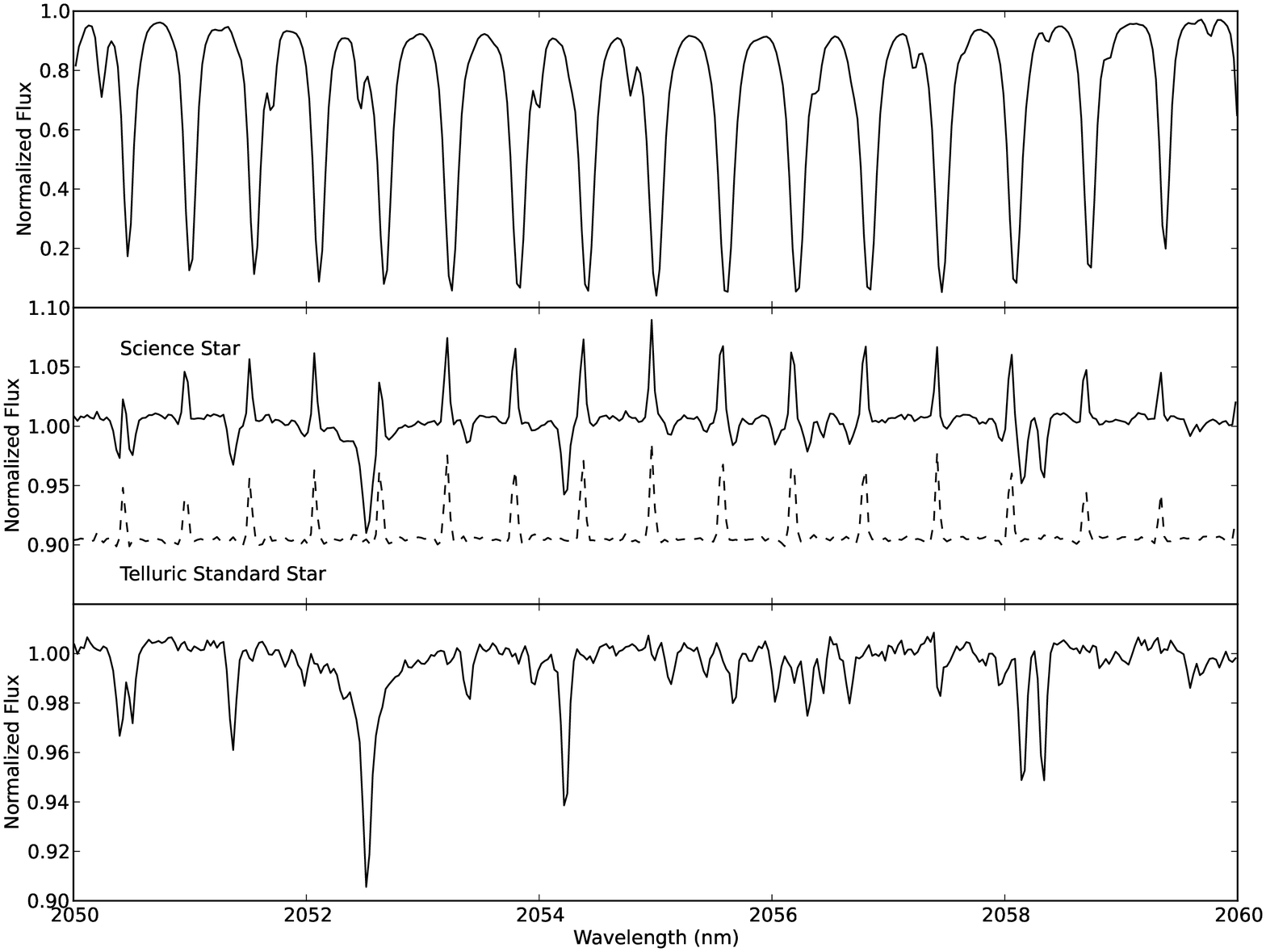}
  \caption{Telluric Correction process. \emph{Top Panel}: Original
    science star spectrum for a segment in the water band on the blue side of the K band.  \emph{Middle Panel}: Telluric residuals
    after the model fit. The science star residuals are in the upper (solid) line, and
    the standard star residuals are in the lower (dashed) line. The systematic errors we introduced appear like emission lines in both spectra. \emph{Bottom Panel}:
    Result after division of the science star residuals by the
    standard star residuals. The telluric contamination has largely been removed.}
  \label{fig:tellcorr}
\end{figure}

Since we are using a telluric model to make the observations, performing a model fit as described above will perfectly remove the telluric spectrum. In order to simulate systematic errors in the model fit, we 
divided both the science star and the standard star by telluric models 
that had $\pm 1\%$ water column density from the ``actual'' value used to make the
observation. This process left large telluric residuals on the order of $5-10\%$ of the continuum, but the residuals were
quite similar in both the science star and the standard star
observations. Thus, division of the science star residuals by the
standard star residuals adequately (but not completely) removes the telluric
contamination. Telluric residuals in the water bands at the edges of the H and K spectral windows were on the order of $1-2\%$ of the continuum. Figure \ref{fig:tellcorr} illustrates the telluric
removal process for a region with particularly severe telluric
contamination.

With the telluric contamination removed, the simulated observations
consist of just the star, the planet, and noise. To generate a stellar spectrum with minimal contamination from either telluric lines or planet lines, we simulate observations of the the star at
various phases in the planet's orbit as well as the Earth's orbit. After
correcting for the Doppler shift from both the Earth's motion and the star's
reflex motion, both of which are assumed known, we co-add the spectra from several observations. This
process will reduce the strength of the planet's spectral lines and any residual telluric 
lines, as well as reduce the random noise in the spectrum by a 
factor of $\sqrt{N}$, where N is the number of observations of the planet. The result is a very high S/N stellar template spectrum, which we subtract from each observation.  While the planetary lines are reduced in intensity in the stellar template,
they are still present at several radial velocities for any finite number of observations. Thus,
subtracting the stellar spectrum will also subtract some of the signal
we are interested in. For observations of the system at N distinct
phases, this stellar subtraction algorithm will subtract 1/N of the
planet signal. Thus, for non-transiting systems we expect the sensitivity to scale approximately linearly with the number of observations at distinct orbital phases. For systems with inefficient heat redistribution, where the different orbital phases contribute different amounts of planet flux, the scaling relation is more complicated and will depend on the phases observed. However, the general result that more observations increase the overall sensitivity is robust.

We note that our method of correcting for the telluric and stellar lines is quite different from that used in most previous work by \cite{Brogi2012, Brogi2013}, and \cite{deKok2013}, although the telluric correction is very similar to that described by \cite{Rodler2012}. In most previous work, the contamination was removed by placing all observations of the star in a matrix and removing features that are stationary in time. This works because the planet will be orbiting, and so its spectrum will shift several pixels throughout the course of the observation while the telluric and stellar lines will remain (approximately) constant. In contrast to this, we perform a physical fit to the telluric spectrum for each individual observation, and we account for the small stellar radial velocity when generating a stellar spectrum template. The method used here is more physical than that of previous work, but can be more expensive in both observational and computational time since it requires the observation of a standard star and the computation of a large number of telluric models over a wide wavelength range.

\subsection{Recovery of Planet Radial Velocity}
\label{sec:rv_recovery}
After removing the telluric and stellar lines, each observation has
been reduced to a very noisy planet spectrum. Unfortunately, the low
planet to star flux ratio of $F_p/F_s \sim 10^{-4}$ means the
random noise generally has an amplitude greater than the variation in
the planet spectrum itself (i.e. the S/N $<1$). The situation is even worse in spectral regions with severe telluric absorption, where small errors in the telluric correction complicate the stellar spectrum removal and effectively add systematic noise. Nonetheless, we can still detect the planet
signature by cross-correlating the residuals against a planet model
spectrum. The cross-correlation will show a peak at the velocity
corresponding to the radial velocity of the planet. 

Except for observations with extremely high S/N ratios, the cross-correlation function (CCF) for a single observation will show several peaks: one for the true planet signal, and several more coming from chance alignments with either random noise or telluric and stellar residuals. Here, we use a method similar to the one described by \cite{Brogi2012, Brogi2013} by using our knowledge of the planet's orbit. For planets found with the radial velocity technique, the \emph{stellar} radial velocity ($v_s$) is known for each observation (or orbital phase $\phi$), and the planet radial velocity ($v_p$) is simply

\begin{equation}
v_p (\phi) = v_s(\phi) \frac{M_s}{M_p}
\label{eqn:planetmass}
\end{equation}

\begin{figure}[t]
  \centering
  \includegraphics[width=\columnwidth]{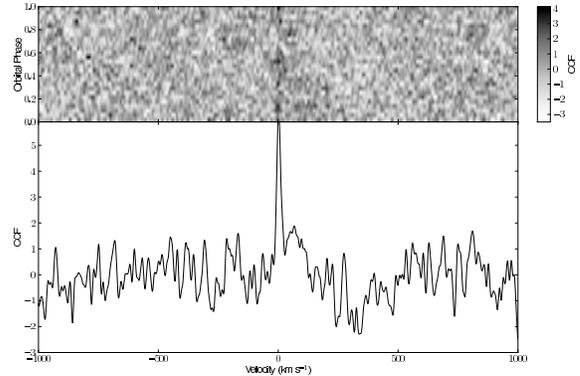}
  \caption{\emph{Top Panel}: Individual cross-correlation functions for simulations of HD 189733 with an average S/N = 500, and with complete heat redistribution. The velocities are shifted at each orbital phase such that the cross-correlation function should have a peak at 0 km s$^{-1}$; the planet is only detected in a few of the observations. \emph{Bottom panel}: The total cross-correlation function, after adding the cross-correlation functions for each individual observation. Here the planet is very clearly detected with $6.29 \sigma$ significance.}
  \label{fig:allcorr}
\end{figure}

To find the true mass of the planet, we test several guess values for the ratio of stellar mass to planet mass ($M_s/M_p$). For each value, we co-add all of the CCFs after correcting for the planet radial velocity and barycentric motion. When the mass-ratio guess is correct, chance alignments with residual noise will tend to cancel out while the CCF peaks coming from the true alignment of the planet model with the planetary spectrum add together. Thus, the total CCF shows a strong peak at 0 km s$^{-1}$, indicating the detection of the planetary spectrum. Figure \ref{fig:allcorr} demonstrates the advantage of adding the cross-correlation functions from all observed orbital phases; even though the planet is only weakly detected in a few of the individual observations, the total CCF has a very strong ($\sim 6 \sigma$) peak. We determine the correct mass-ratio by comparing the height of the CCF at 0 km s$^{-1}$ for all of the mass-ratio guesses; the correct mass-ratio will have the highest CCF peak.

\section{Results}
\label{sec:results}
\begin{figure}[t]
  \centering
  \includegraphics[width=\columnwidth]{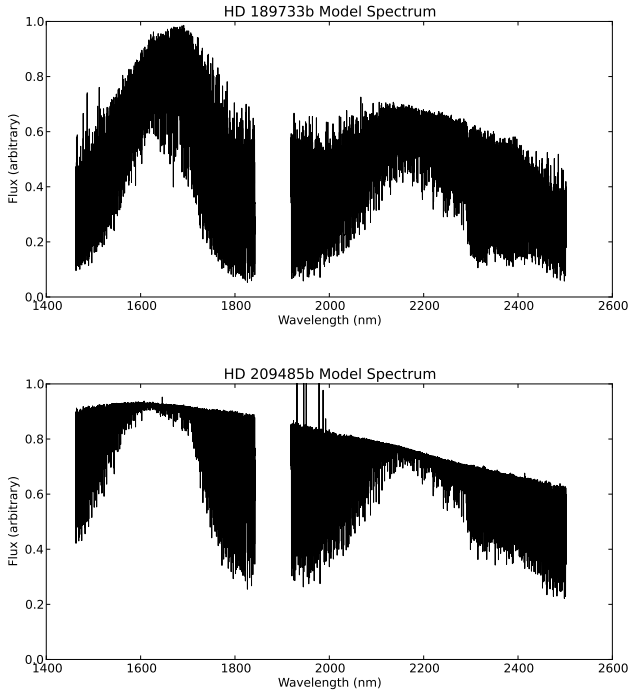}
  \caption{A comparison of the HD189733 planet model with the HD209458
  planet model. The gap in the middle is is in between the H and K
  spectral windows, where water absorption in the Earth's atmosphere
  blocks most of the incoming light. HD209458b has a thermal inversion
  in its atmosphere, which generates the emission lines near 1950
  nm. HD189733 b is somewhat cooler and has more water, giving its
  spectrum stronger spectral lines.}
  \label{fig:modelcomp}
\end{figure}

We simulated a series of IGRINS observations of non-transiting Hot Jupiter systems by using model spectra for the well studied systems HD189733 and HD209458. For both systems, we calculate the minimum S/N ratio necessary to detect the planet. In this work we consider a planet detected if the total (summed) CCF has a peak at 0 km s$^{-1}$ with at least $4\sigma$ significance, and that the planet mass is recovered correctly and unambiguously. We consider the effect of heat redistribution by taking two extreme cases as described in Section \ref{sec:obsgen}, and we estimate the model dependence of our method by cross-correlating our synthetic observations against the wrong planet model spectrum. Our method of testing the model dependence is most likely pessimistic; in a real observing campaign and indeed in current searches \citep{Brogi2012, Brogi2013, Rodler2012, deKok2013} a library of planetary atmosphere grids with different temperature, pressure, and molecular abundance profiles would be tested. We consider it likely that one model in such a model library would be closer to the true planet spectrum than our two test cases are to each other (see Figure \ref{fig:modelcomp} for a visual comparison of the planet model spectra).

In general, the critical S/N ratio necessary to detect any non-transiting planet with our method depends on the ratio of the signal coming from the planet to that of the star. This ratio is not simply the flux ratio, since our cross-correlation technique requires several deep spectral lines in the planetary spectrum. If the planetary atmosphere is nearly isothermal as in WASP-12b \citep{Crossfield2012}, or has a thick haze layer in the near-infrared as HD 187333b may \citep{Gibson2012}, then the spectrum will be relatively featureless and it will be very difficult to detect in high resolution.  The critical S/N ratio can be calculated from the continuum surface flux of the planet and star ($F_{\rm p, cont}$ and $F_{\rm s, cont}$, respectively), the ratio of the flux in an average spectral line to the flux in the continuum for the planet ($r$), the radii of the planet and the star ($R_p$ and $R_s$, respectively), and unknown constants $A$ and $S_0$:

\begin{equation}
S/N_{\rm crit} = A \frac {F_{\rm s, cont}}{F_{\rm p, cont}(1 - r)} \left ( \frac{R_s}{R_p} \right )^2 + S_0
\label{eqn:snrcrit}
\end{equation}

To first order, the continuum fluxes can be calculated from the blackbody fluxes and the effective temperatures of the planet and star. The planet temperature can be approximated from the stellar temperature, the semimajor axis of the planet's orbit ($a$), and the bond albedo of the planet ($A_B$) by radiative equilibrium with

\begin{equation}
T_p^4 = \frac{1-A_B}{4} \left ( \frac{R_s}{a} \right )^2 T_s^4
\end{equation}

To test the dependence on S/N, we generated a series of synthetic observations at 25 approximately evenly spaced orbital phases for average S/N ratios ranging from 100  - 1500, with both efficient and inefficient heat redistribution (see the discussion of heat redistribution in section \ref{sec:obsgen}). We did not attempt to optimize the observing schedule for the planets with inefficient heat redistribution, and so the minimum S/N ratios we find for that case may be somewhat pessimistic. We determined the minimum S/N ratio necessary to detect the planet, and report the results in Table \ref{tab:snrcrit}. In general, HD 209458b requires higher S/N to detect than HD 189733b. HD 209458 is a hotter star while the planets are roughly the same temperature, and so the flux ratio is more extreme. In addition, HD 189733b has much deeper lines throughout its spectrum, largely from a higher water abundance \citep{Madhusudhan2009}, making the detection easier. A second trend evident in Table \ref{tab:snrcrit} is that planets with inefficient heat redistribution require a larger increase in S/N ratio to detect if the wrong planet model is used, and so are more model dependent than those with efficient heat redistribution.

\begin{table*}[t]
  \centering
  \begin{tabular}{|c|ccc|}
    \hline
    Star & Planet Model for CCF & Heat Redistribution & Critical S/N Ratio \\ \hline
    HD 189733 & HD 189733b & efficient & 200 \\
    HD 189733 & HD 189733b & inefficient & 450 \\
    HD 189733 & HD 209458b & efficient & 300 \\
    HD 189733 & HD 209458b & inefficient & 800 \\
    HD 209458 & HD 209458b & efficient & 900 \\
    HD 209458 & HD 209458b & inefficient & 1200 \\
    HD 209458 & HD 189733b & efficient & 1200 \\
    HD 209458 & HD 189733b & inefficient & $>$1500 \\ \hline
  \end{tabular}
  \caption{Summary of minimum S/N ratios needed to detect planets for our test cases. Efficient heat redistribution refers to observations where the dayside and nightside temperatures are the same, while inefficient heat redistribution is when the nightside temperature is half that of the dayside temperature. The model dependence is estimated by using different planet models to cross-correlate against the same observation (see section \ref{sec:results}). We do not attempt to simulate observations with S/N $>1500$ because such a high value would be very difficult to achieve in the near-infrared.}
  \label{tab:snrcrit}
\end{table*}

We now determine $A$ and $S_0$ from Equation \ref{eqn:snrcrit} from the known flux ratios and line strengths in HD 189733b and HD 209458b, along with the minimum S/N ratio values in Table \ref{tab:snrcrit}. For efficient heat redistribution, $A=0.1$ and $S_0 = 46$. Likewise, $A=0.1$ and $S_0 = 107$ for inefficient heat redistribution. With these values, we can estimate the S/N ratio necessary to detect other Hot Jupiters. The planet radius is not known for non-transiting planets, but we approximate this from mass-radius relationships for hydrogen-dominated planets given in \cite{Swift2012}. Lacking any information on the line strength for non-transiting planets, we take a typical value to be the average of our two test cases: $r = 0.83$. We use $A=0.1$ and $S_0=75$ in Equation \ref{eqn:snrcrit} to consider some level of heat redistribution. Table \ref{tab:targetlist} estimates the flux ratio for each of the non-transiting Hot Jupiter planets in the exoplanets.org database \citep{exoplanet}, assuming $A_B = 0.2$. The minimum S/N ratios were found with Equation \ref{eqn:snrcrit}. We calculate the exposure time for each target from the net atmospheric transmission in the K band, the expected $7\%$ net throughput of IGRINS (Dan Jaffe, priv. comm.), and simple Poisson statistics since these observations will be well within the source noise limit.

\begin{figure}[ht]
  \centering
  \includegraphics[width=\columnwidth]{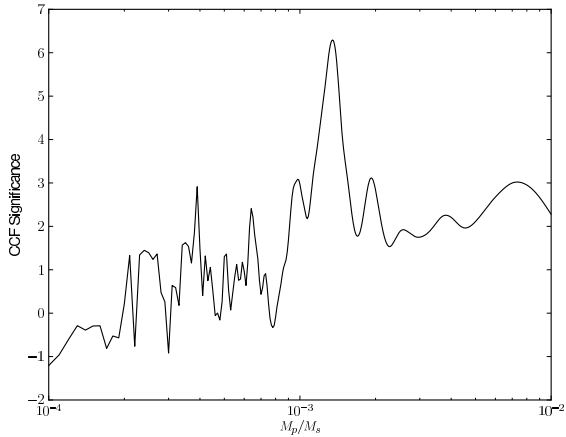}
  \caption{A summary of the significance of the point at 0 km s$^{-1}$ in the total cross-correlation functions for each of the mass-ratio guesses, for a series of observations of HD 189733 with an average S/N = 500 and complete heat redistribution (the same as in Figure \ref{fig:allcorr}). The peak in this figure determines the planet to star mass-ratio, and therefore the true planet mass. The width of the peak $1 \sigma$ below its maximum determines the uncertainty in the planet mass.}
  \label{fig:massratio}
\end{figure}

To estimate the uncertainty in the planet to star mass-ratio, and therefore the uncertainty in the true planet mass, we compare the significance of the $v=0$ km s$^{-1}$ point of the total CCFs for various mass-ratio guesses. As the guess begins getting closer to correct, the correct peaks in the individual CCFs will begin to align and the significance of the total peak will increase (See Figure \ref{fig:massratio}). We can estimate the mass-ratio uncertainty from the points where the significance of the total CCF peak drops $\sim 1 \sigma$ from the most significant mass-ratio. Smaller mass-ratio systems will have faster moving planets, so the individual CCF peaks only line up for a more narrow range of mass-ratio guesses than they would for more massive planets. Therefore, the the planet mass uncertainty with this method \emph{decreases} as the mass-ratio decreases, as long as the planet remains detectable. The width of the peak in Figure \ref{fig:massratio}, which is for a detection of HD 189733b, is $\sigma_{M_p/M_s} = 1.6\e{-4}$. Combining this with the stellar mass uncertainty gives a planet mass uncertainty of $\sigma_{M_p} = 0.15 M_{\rm Jup}$. The width of the corresponding peak for a detection of HD 209458b is $\sigma_{M_p/M_s} = 2\e{-5}$ giving $\sigma_{M_p} = 0.03 M_{\rm Jup}$. The planet mass uncertainty could be improved by focusing on phases near quadrature, rather than evenly sampling the orbit as we do in this paper.

\section{Summary and Conclusions}
\label{sec:summary}
We have described a technique for directly detecting the near-infrared spectrum from a 
non-transiting Hot Jupiter using high resolution, high signal-to-noise ratio spectra. We have applied this technique to simulated observations with the IGRINS near-infrared instrument, which is sensitive to the entire H and K bands and will begin operating on the 2.7m Harlan J. Smith Telescope at McDonald Observatory in late 2013. Using models of the well-studied Hot Jupiters HD189733b and HD209458b, we make synthetic observations at various phases of the planet's orbit and at several different barycentric radial velocities to simulate observations taken at different times of the (Earth's) year. We then remove the telluric absorption and parent star spectra, and search for the planet spectra by cross-correlating the residuals against models of the planet spectrum.

We have shown that the true mass and inclination of a Hot Jupiter planet can be recovered, and have determined the effect of S/N ratio as well as heat redistribution, and have estimated the model dependence of our method. Table \ref{tab:snrcrit} summarizes the results for our two test case planets, and Table \ref{tab:targetlist} estimates the S/N ratios and exposure times necessary to detect several known, non-transiting Hot Jupiters.

Our simulated observations are similar to the recent detections of Tau Boo b \citep{Rodler2012, Brogi2012}, HD 189733 b \citep{deKok2013}, and 51 Peg b \citep{Brogi2013}, all of which used the CRIRES high resolution spectrograph. The IGRINS instrument covers the entire H and K bands at once, rather than the $\sim 40$ nm range observed by CRIRES. Since the cross-correlation signal roughly scales as the square root of the number of deep spectral lines, we expect IGRINS to be more sensitive to detecting planets than CRIRES despite being on a smaller telescope. As well as simulating a different instrument, we simulate a different observing strategy; whereas previous work has used several hundred observations of a star over the course of a few closely-spaced nights, we have simulated an observing strategy where the star is observed $\sim 25$ times at various points in its phase \emph{as well as various times of the year}. This strategy is more compatible with a campaign to monitor several Hot Jupiters rather than using one observing run for each planet. It also allows for the easy addition of more data when it is received.

This research has made use of the Exoplanet Orbit Database and the Exoplanet Data Explorer at exoplanets.org. We would like to thank the anonymous referee for several very helpful comments, Travis Barman for generating the high-resolution model spectra for the stars and planets used in this work, and Dan Jaffe for his help with estimating the performance of the IGRINS instrument.

\begin{table}[ht]
  \centering
  \begin{tabular}{|ccccc|}
    \hline
    Star Name & $ \rm K_s$& $\rm K_s$-band & S/N & Exposure Time \\
    & Magnitude & flux ratio & required & (minutes) \\ \hline
    $\tau$ Boo  & 3.36 & 1.26\e{-03} & 541 & 1.0 \\
    HD 189733  & 5.54 & 3.08\e{-03} & 200 & 1.0 \\
    $\upsilon$ And  & 2.86 & 8.22\e{-04} & 791 & 1.4 \\
    HD 41004 B  & 6.43 & 8.95\e{-03} & 141 & 1.7 \\
    HD 179949  & 4.93 & 1.22\e{-03} & 559 & 4.7\\
    HD 162020  & 6.54 & 2.99\e{-03} & 272 & 4.8 \\
    HD 217107  & 4.53 & 7.22\e{-04} & 890 & 8.2 \\
    HD 73256  & 6.26 & 1.53\e{-03} & 461 & 10.8 \\
    HD 187123  & 6.34 & 1.28\e{-03} & 534 & 15.5 \\
    HD 86081  & 7.3 & 1.64\e{-03} & 434 & 25.0 \\
    HD 68988  & 6.74 & 1.17\e{-03} & 576 & 26.3 \\
    HIP 14810 & 6.83 & 1.05\e{-03} & 637 & 34.8 \\
    HD 330075  & 7.17 & 1.23\e{-03} & 553 & 35.9 \\
    HD 149143  & 6.43 & 7.71\e{-04} & 838 & 41.9 \\
    HD 209458  & 6.31 & 1.21\e{-03} & 900 & 43.0 \\
    HD 185269  & 5.26 & 3.05\e{-04} & 2002 & 81.0 \\
    HD 118203  & 6.54 & 4.77\e{-04} & 1309 & 112.9 \\
    HD 102956  & 5.66 & 2.06\e{-04} & 2935 & 253.0 \\ \hline
  \end{tabular}
  \caption{Estimated exposure times required for correctly retrieving the true masses of Hot Jupiters. The times are for a single observation; roughly 20-25 observations of the system at different phases are necessary to detect the planet. See Section \ref{sec:results} for the calculation of the flux ratio and critical S/N.}
  \label{tab:targetlist}
\end{table}

\newpage 

\end{document}